\documentclass[preprint, aps, prc, 10pt, nofootinbib, superscriptaddress, floatfix]{revtex4-1}
\usepackage{anyfontsize}
\usepackage{hyperref}
\usepackage{natbib} 
\usepackage{soul}
\usepackage{multirow}
\usepackage{tabularx}
\usepackage{amsthm}
\usepackage{amsfonts}
\newtheorem{definition}{Definition}

\usepackage[T1]{fontenc}
\usepackage{xcolor}
\usepackage{graphicx} 
\graphicspath{{figures/}}
\usepackage{dcolumn} 
\usepackage{bm} 
\usepackage{multirow}
\usepackage{placeins}

\usepackage{tabularx} 

\newcommand\clearrow{\global\let\rowmac\relax}
\newcommand{\iso}[2]{${}^{#2}$#1}

\usepackage{etoolbox}
\apptocmd{\sloppy}{\hbadness 10000\relax}{}{}

\def\kr83{{${}^{83}$Kr${}^{\mathrm{m}}$}}

\begin{document}

\title{Physics-driven Comparative Analysis of Various Statistical Distance Metrics and Normalizing Functions}

\newcommand{\iu}{Center for Exploration of Energy and Matter, Indiana University, Bloomington, IN 47405, USA}

\author{N.~Fuad}
\affiliation{\iu}

\begin{abstract}
    Comparison of two probability density/mass functions (PDF/PMFs) is ubiquitous in various forms of scientific analysis, including machine learning, optimization problems, and hypothesis tests. A copious amount of distance metrics have already been proposed and are regularly being used in this regard. In this document, we report a data-driven systematic comparison among a few of such metrics. The metrics considered here are Hellinger distance, Wasserstein distances (1D), $\sqrt{JS}$ distance, $L_\infty$ norm, Kolmogorov-Smirnov distance, and Fisher-Rao metric. We perform this comparison using electron and photon events from a decaying \iso{Kr}{83} isotope, collected through an HPGe spectrometer operating under cryo-vacuum conditions. To accomplish this, first, a dimensionless Parameter of Interest (PoI) was established, then PDF/PMFs were generated from the data, and finally the stabilities of the PoI under various criteria, such as sample size, discretization length, and normalizing functions, were studied and the results were summarized. In this report, we also propose a list of properties that a normalizing function should have and utilize them in the comparison.
\end{abstract}
\maketitle

\section{Introduction}
Necessity of a standardized measurement of dissimilarity between Probability Density/Mass Functions (PDF/PMFs) is ubiquitous across many fields of scientific analysis. A copious amount of parameters to do so have already been introduced and are regularly used \cite{Aler2020, Arjovsky2017, Kim2004, lin1991divergence, endres2003new}. Many of these parameters have also been shown to be connected to each other (e.g. Hellinger distance can be written in terms of Bhattacharya coefficient \cite{kailath1967divergence, Jolad2016} or Chernoff divergence \cite{nielsen2022revisiting}, Fisher information metric is proportional to $\sqrt{}$JS \cite{CasasLambertiPlastino2004, Edelsbrunner2019, sra2019metrics, Nielsen2020} , Wasserstein-2 takes shape of $\chi$-distance
in 1-D \cite{Ramdas2015, Tang2023}). Commonly used such parameters can be broadly put into two categories - \textit{metrics} and \textit{non-metrics}. They are sometimes referred to as \textit{distances} and \textit{divergences} respectively also [e.g. \cite{sra2019metrics}]. A function $d(x,y)$ is called a metric if it satisfies the following properties \cite{Bruno2014} -  
    \begin{itemize}
   \item Non-negativity : $d(x,y)\geq 0$
   \item Identity : $d(x,x)=0$
   \item Symmetry : $d(x,y)=d(y,x)$
   \item Triangle inequality : $d(x,y)+d(y,z) \geq d(x,z)$
\end{itemize}

\newcommand{\sumover}{\sum}

\newcommand{\limfrac}[2]{\genfrac{}{}{0pt}{}{\,\lim\,}{#1 \rightarrow #2}}

In this analysis, we compare following \textit{metrics} between two discretized PDF/PMFs, $p(x)$ and $q(x)$  (List \ref{list:equations})- 
\begin{enumerate}
\label{list:equations}
    \item Hellinger distance \cite{Aler2020}
    \begin{equation}
        H(p,q)=\sqrt{\sumover \frac{1}{2} \left( \sqrt{p} - \sqrt{q} \right)^2 }
    \end{equation}
    \item Wasserstein-1 distance \cite{weng2019, wasserstein_scipy, Ramdas2015}
    \begin{equation}
        W_1(p,q)=\sumover |F_{p}^{-1} - F_{q}^{-1}|
    \end{equation}
    where, $F_p^{-1}$ and $F_q^{-1}$ are the quantile functions of $p$ and $q$ 
    \item Wasserstein-2 distance \cite{Tang2023}
    \begin{equation}
        W_2(p,q)=\sqrt{\sumover \left( F_{p}^{-1} - F_{q}^{-1} \right)^2}
    \end{equation}
    where, $F_p^{-1}$ and $F_q^{-1}$ are the quantile functions of $p$ and $q$ 
    \item $\sqrt{}$Jensen-Shannon \cite{endres2003new,CasasLambertiPlastino2004}
    \begin{equation}
        \sqrt{JS(p,q)}= \sqrt{\sumover \frac{1}{2}\left(p\log \frac{2p}{p+q}+q\log\frac{2q}{p+q}\right)}
    \end{equation}
    \item $L_\infty$ norm/Chebyshev distance
    \begin{equation}
        L_\infty(p,q)=\lim_{r\rightarrow\infty} \left(\sumover|p-q|^r\right)^\frac{1}{r}=\max(|p-q|)
        \end{equation}
    \item Kolmogorov-Smirnov distance
    \begin{equation}
        KS(p,q)=\max(F_p-F_q)
    \end{equation}
    where, $F_p$ and $F_q$ are the CDFs of $p$ and $q$
    \item Fisher-Rao distance 
    
Closed-form expression of Fisher-Rao distance for a categorical distribution was used. A few other closed-form results can be found in \cite{Miyamoto2024}.
    \begin{equation} 
        FR(p,q) = \frac{2}{\pi}\arccos\left(\sumover\sqrt{pq}\right)
    \end{equation}
\end{enumerate}
where, $p(x)$ and $q(x)$ in are either PDF or PMF in the usual sense - $p(x)$ and $p_m(x)$ are PDF and PMF respectively on the parameter space $\chi={\{x\}}$ if they satisfy following properties -  
\begin{itemize}
    \item $p(x)\geq0; 0\leq p_m(x)\leq1 \forall x\in\chi$
    \item $\sumover p_m(x)=\sumover p(x)dx=1$
\end{itemize}
\newcommand{\ra}{\rightarrow}

Since $p(x)$ is unbounded from upper side, distance measurement using $p(x)$ can not be guaranteed to be upper bounded. But $p_m(x)$ are bounded in $[0,1)$ by construction. We utilize this notion of normalization to as an additional comparison of our proposed normalized functions. We also present another measure of goodness of normalization which is independent of this. Metrics, such as Fisher-Rao (FR), which include $\arccos$ are bounded by \texttt{domain}:$[0,1]$; to tackle these cases, normalizing function is applied right after summation is applied, which is essentially $\sum\rightarrow n(\sum)$ or $\max\ra\ n(\max)$. Here we propose a list of properties that a normalizing function should have - 

\begin{definition}
\label{def:norm_funcs}
    We define a function, $n(x)$, to be a normalizing function if $n(x)$ satisfies the following properties - 
    \begin{itemize}
        \item Bounds: $\displaystyle \lim_{x\ra0} n(x)=0$ and $\displaystyle \lim_{x\ra\infty} n(x)=1$; \\
        $\mathrm{Domain:[0,\infty)}, \mathrm{Range:[0,1)}$
        \item Bijectivity: Ensures existence of an inverse function, $n^{-1}(x)$
        \item Inverse bounds: $\displaystyle \lim_{x\ra0} n^{-1}(x)=0$ 
        and $\displaystyle \lim_{x\ra1} n^{-1}(x)=\infty$;
        $\mathrm{Domain^{-1}:[0,1)}$ and $\mathrm{Range^{-1}:[0,\infty)}$.
        \item Monotonicity: $x_1>x_2 \Rightarrow n(x_1)>n(x_2)$ 
        \item Metric preservation: If $d(x,y)$ is a metric, then $n\circ d$ must also be a metric \cite{DragomirNicas2018}
    \end{itemize}
    
    Note 1. If $d(x,y)$ is a metric, and $n(x)$ is monotonic, then $n\circ d$ is guaranteed to be a metric iff $n(0)\geq0$ and $n(x)$ is concave \cite{DragomirNicas2018,Niculescu2025}.
    
    Note 2. If $n\circ d$ is metric preserving through concavity of $n(x)$, then $n^{-1}(x)$ will necessarily be convex and $n^{-1}\circ d$ will not be metric preserving.
\end{definition}

In this analysis, we consider the following list of functions that follow Def. \ref{def:norm_funcs} - 
\begin{itemize}
    \item $n_1(x)=\frac{\log(1+x)}{1+\log(1+x)}$
    \item $n_2(x)=\frac{x}{1+x}$
    \item $n_3(x)=1-e^{-x}$
    \item $n_4(x)=\frac{2}{\pi}\arctan(x)$
\end{itemize}

\textit{Notes}
\begin{enumerate}
    \item True domains and ranges of the $n(x)$s are supersets of domains and ranges defined in Def. \ref{def:norm_funcs}, not equal; e.g. true domain of $n_1(x)$ is $(-1,\infty)$ and of $n_2(x)$ is $(-\infty,\infty)\setminus\{-1\}$, not $[0,\infty)$.
\end{enumerate}

\begin{figure}[htbp]
    \centering
    \includegraphics[width=0.9\linewidth]{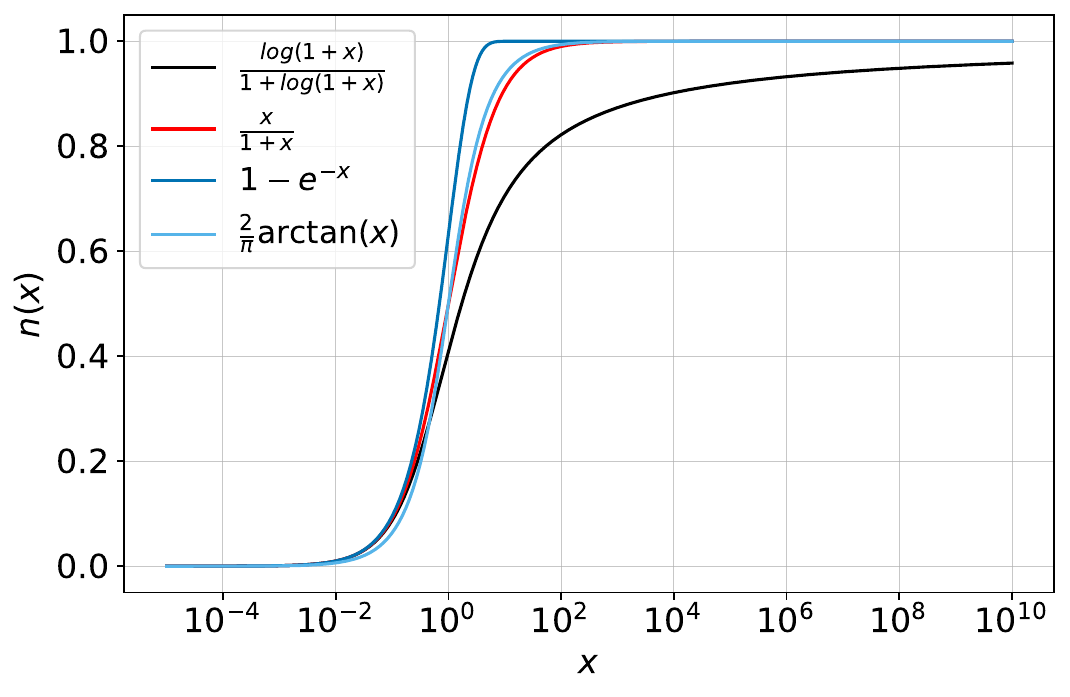}
    \caption{$n(x)\forall x: x\in \{-5\leq \log(x) \leq 10\}$. It can be observed that $n_{2-4}(x)$ saturate much faster compared to $n_1(x)$ and reaches $1$ within $x=\mathcal{O}(10^2)$. So any distance metric that produces value $>>10^2$ will not be distinguished by $n_{2-4}(x)$. In the unsaturated region, $n(x)$s behave distinctively until it reaches $\mathcal{O}(10^{-2})$}
    \label{fig:norm_func}
\end{figure}

\FloatBarrier

\section{PDF/PMF Generation}
\label{methodology}
\begin{figure}[!htbp]
        \includegraphics[width=0.9\columnwidth]{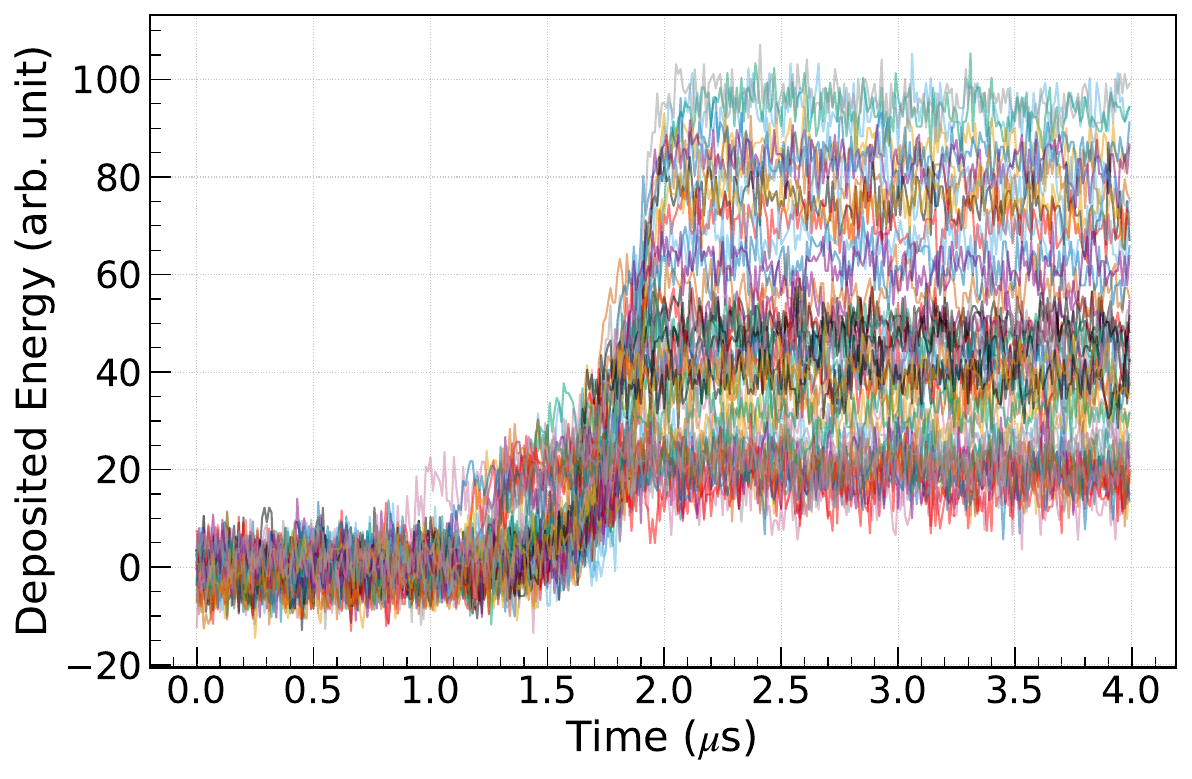}
        \caption{Examples of signals in HPGe detectors. Each signal consists of samples taken at $10$ ns intervals over a $0.2$--$2~\mu$s window. Signal values from well before to well after the trigger ($\mathcal{O}(10~\mu\mathrm{s})$) are also recorded to analyze the quality of the signal. The height of the signals corresponds to the amount of energy deposited.}
        \label{fig:waveforms}
\end{figure}

The data were collected using a High Purity Ge based spectrometer exposed to an \iso{Kr}{83} source. The detector used in our study is known as PPC type detector \cite{Edzards2021, luke1988low} due to its geometry and type of charge collection system. Ge is a semiconductor, so when some energy deposition takes place inside the detector, it creates a number of electron-hole pairs which is dependent on deposition energy, operating temperature, geometry of the detector leading to different amount of energy loss through other mechanism such as ionization, phonon creation or recombination etc. These charges are collected at a point on the surface by applying an electric field through the detector, typically in $\mathcal{O}(\mathrm{kV})$. These charges change the voltage of the circuit, which is written to disk using an ADC and converted to a parameter with dimension of energy based on the effective capacitance of the detector. The clocking resolution of our apparatus was $10$ ns, meaning the above mentioned parameter was estimated every $10$ ns. A set of values of this parameter around the time of energy deposition constitute our signals, $s(t)$, which are called \textit{waveforms} \cite{Abgrall_2022}. Fig. \ref{fig:waveforms} shows a randomly selected set of such waveforms from one of our data-taking campaigns. These data were taken at cryogenic vacuum pressure ($\mathcal{O}(\mathrm{nbar})$) and Liquid Nitrogen temperature ($\sim 88$K).

\begin{figure}[!htbp]
    \centering
    \includegraphics[width=0.85\linewidth]{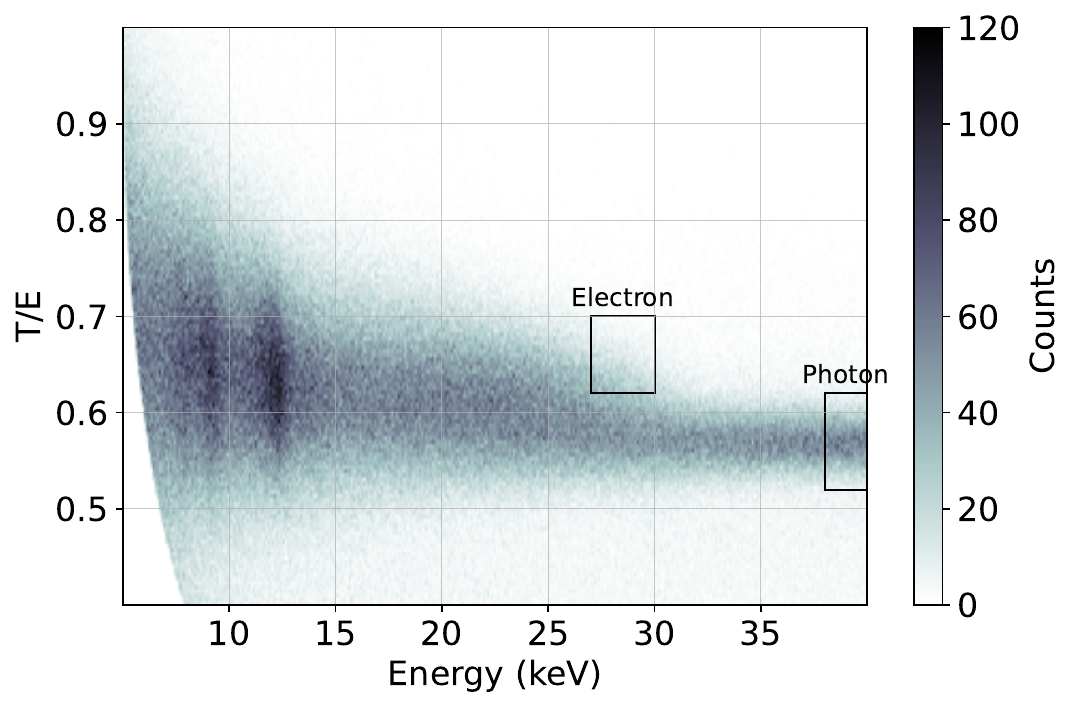}
    \caption{Spectrum of \iso{Kr}{83} on $T/E$ parameter space. It excludes the events cut based on \texttt{stability}, \texttt{drifttime} and \texttt{$t_0$}. The selected events are shown in black colored boxes. There were $6,982$ electron events and $14,146$ $\gamma$ events. These events did not have any cut based on $A/E$.}
    \label{fig:kr_spec_toe}
\end{figure}

The energy region of the daughter particles of \iso{Kr}{83} fall at or below $\sim32$ keV \cite{kr_nudat}. There are also photon events below this energy from \iso{Kr}{83}, but due to a well known property of the HPGe detectors' response to charged particles \cite{Gruszko2022}, photon events are mixed together with the electron events near the photon deposition energies. This makes these photons unusable in our case. But knowing that the upper limit of the electron events from \iso{Kr}{83} is $\sim32$ keV allows us to use the Compton photons, of $38$--$40$ keV, which are ubiquitous between $\mathcal{O}(\mathrm{keV})$ and $\mathcal{O}(\mathrm{GeV})$. We utilize two parameter spaces of HPGe detectors known as $T/E$ and $A/E$. $T$ was measured by convolving a triangle filter to each waveform and taking maximum of it, which makes it essentially $s(t)_{max}$. $E$ was measured by convolving a trapezoidal filter \cite{Tan2004}. We use $T/E$ to isolate the two populations of events (photons and electrons) and $A/E$ to build the probability functions on. More details and its use in the MAJORANA DEMONSTRATOR can be found in Ref. \cite{Alvis2019multisite}. Here we use a simpler version of this parameter, $A_{max}/E$. Fig. \ref{fig:kr_spec_toe} shows all electron and photon events used in this analysis between $5$--$40$ keV on a $T/E$ space. 
The following considerations were made during the event selection process - 
\begin{itemize}
    \item \textit{Detector stability:} Events that were collected when the detector was not stable were discarded based on multiple detector condition parameters. For example, if the detector readout voltage was not stable for at least $\mathcal{O}(10\mu s)$ before event triggers or was stable but not flat, those events were discarded. 
    \item \textit{Drifttime:} The time between start and end of charge collection for any event in the detector used are expected to be in $\mathcal{O}(1)$ $\mu s$. Any event having $\geq10 \mu s$ are considered anomalous in this analysis and discarded.
    \item \textit{Error in $t_0$ estimation:} The estimation of the start of charge collection time, $t_0$, is susceptible to noise, so a $3\sigma$ cut was applied on this parameter.
\end{itemize}

\begin{figure}[htbp]
    \centering
    \includegraphics[width=0.85\linewidth]{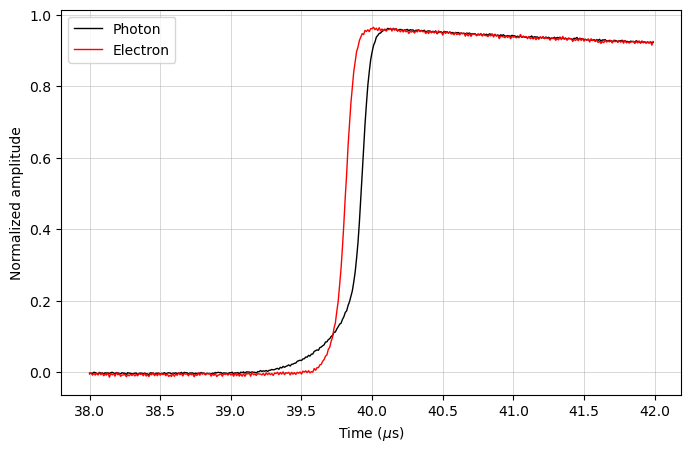}
    \caption{Averaged waveforms from populations of two types of signal events considered in this analysis. The distinctiveness of these two types of events can be observed in the rising edges of these two waveforms.}
    \label{fig:e_vs_gamma_superpulse}
\end{figure}
Fig. \ref{fig:e_vs_gamma_superpulse} shows the \texttt{superpulses} of each event type. These \texttt{superpulses} are created by taking direct average of $1000$ waveforms ($s(t)$) of each event type.  We can observe from Fig. \ref{fig:e_vs_gamma_superpulse} that electron signals rise much more sharply than the photon signals. It is the defining distinction between electron and photon signals for our analysis. The reason for this is as follows -  electrons are charged particles whereas photons are neutral, so when an electron enters the detector, it gets stopped very early in its trajectory inside the detector ($\mathcal{O}(\mu m)$ compared to photon events ($\mathcal{O}(100 \mu m)$). So the signal generated by the electrons move to the charge collection point (which sits on the surface of the detector) much faster than photon signals and it shows up in the sharpness of their rising edges.

\begin{figure}[!htbp]
    \centering
    \includegraphics[width=0.9\linewidth]{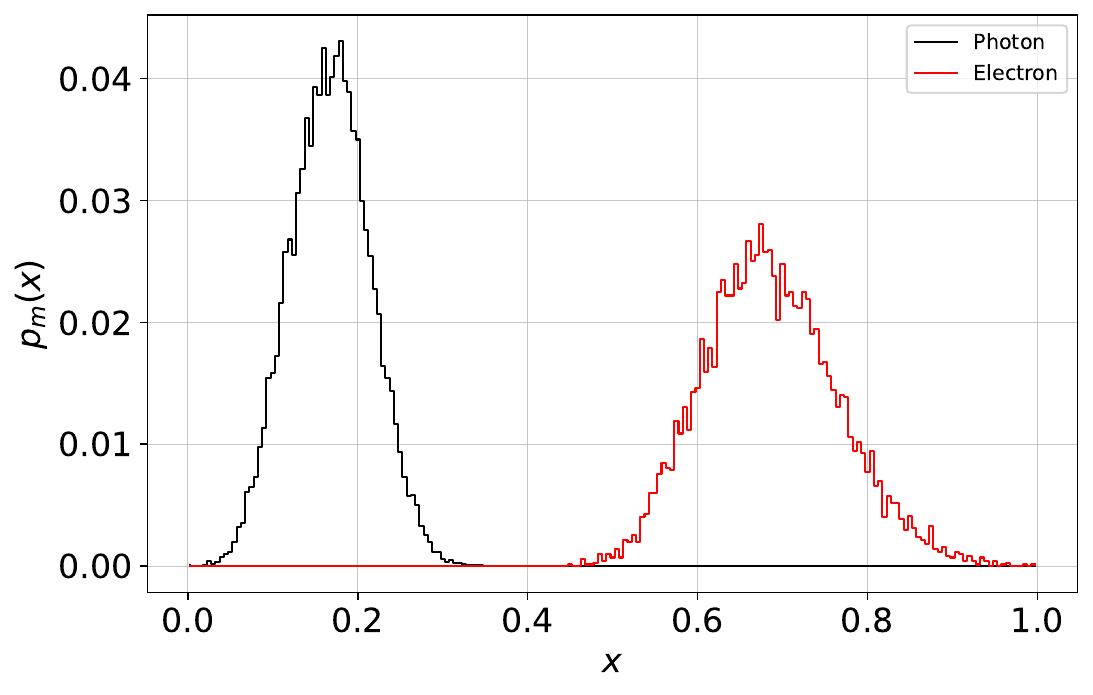}
    \caption{PMFs of the selected electron and photon events. It can be visually noticed the two distributions are disjoint, but not maximally disjoint. We use this observation to make conclusions in later sections.}
    \label{fig:pmf}
\end{figure}

Many different parameters are possible to quantify the sharpness of the rising edge of the waveforms. In this analysis we use $x=\max \left(\frac{ds(t)}{dt}/E\right)$. This parameter has the dimension of energy. This was then rescaled by $x:\rightarrow x=\frac{x-\min(P\cup Q)}{\max(P\cup Q)-min(P \cup Q)}$, where $P$ and $Q$ are the sets of all observed values of $x$ for electron and photon events. This rescaled $x:x \in \chi \subseteq [0,1]$ is dimensionless and used as the parameter of interest (POI). PMFs (discretized) generated for electron and photon events from their respective values of $x$ are shown in Fig. \ref{fig:pmf}. Lower values of $x$ correspond to slower rises, so photons have lower $x$ values, where electrons have higher in this parameter space.

\FloatBarrier

\section{Results}

\begin{figure}[!htbp]
    \centering
    \includegraphics[width=0.9\linewidth]{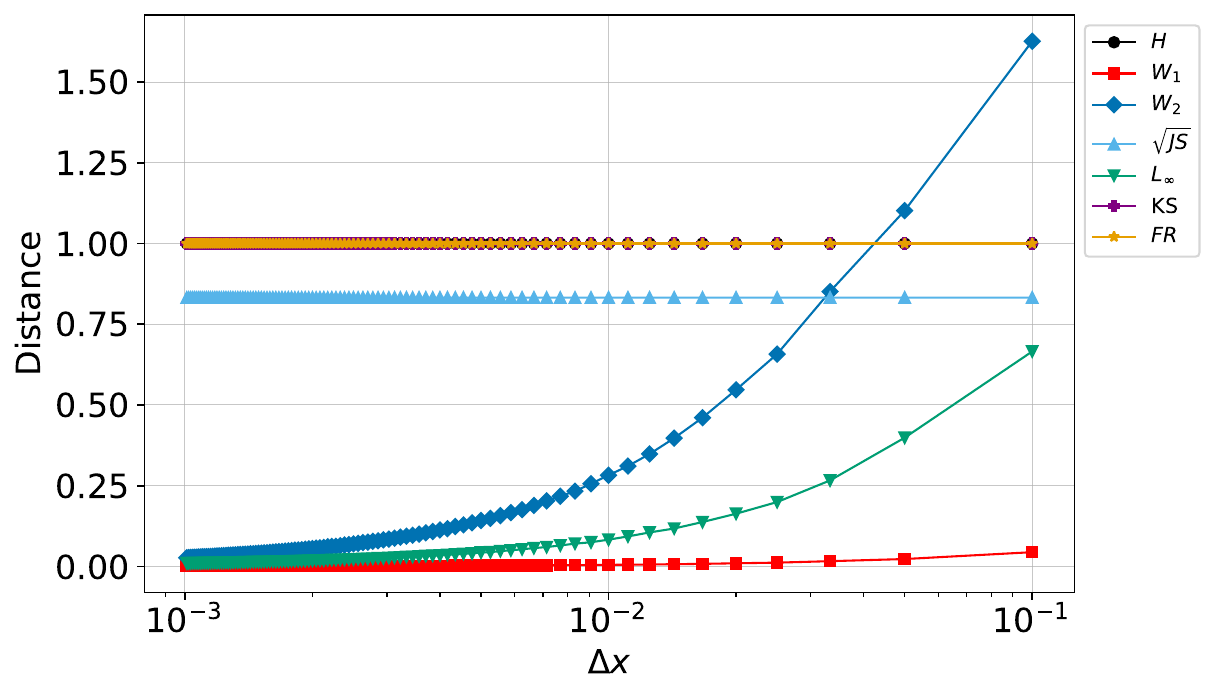}
    \caption{Discretization length (unitless) is varied to investigate the stability of \textit{distances} w.r.t. this change. These distances are $n_0(d(p,q))$. $W_2$ is impacted by this the most whereas Hellinger, $\sqrt{}JS$ and Fisher-Rao distances are observed to be stable.}
    \label{fig:dist_vs_binsize}
\end{figure}

\begin{figure}[!htbp]
    \centering
    \includegraphics[width=0.7\linewidth]{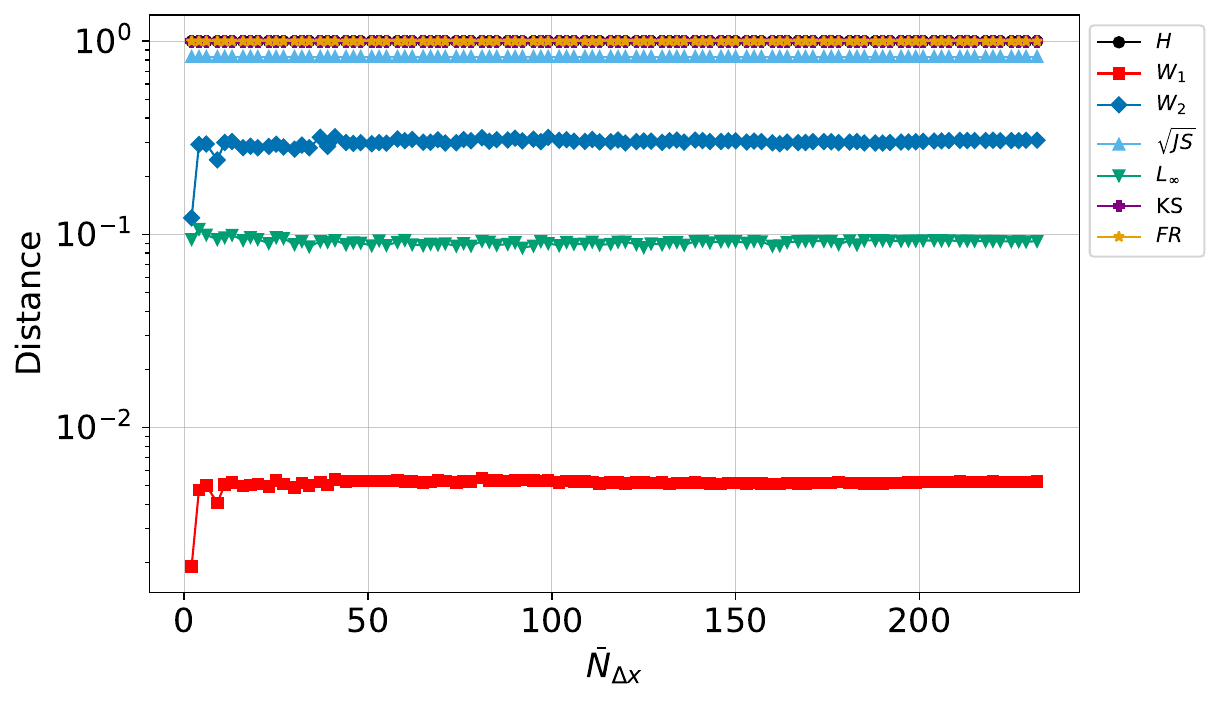}
    \caption{Effect of sample sizes used to generate the PDF/PMFs to the \textit{distances} measured by each metric. These distances are also $n_0(d(p,q))$. $\bar{N}_{\Delta x}$ refers to average number of events within the discretization length, which is taken to be $0.01$. $\sqrt{JS},L_\infty$ and $FR$ \textit{distances} are observed to be most unstable under low statistics.}
    \label{fig:dist_vs_events_frac}
\end{figure}

Tab. \ref{tab:distance_metrics} tabulates the distances measured by the metrics described in List \ref{list:equations}. In case of $n_0(x)$, the values refer to \textit{distances} between PMFs whereas in other cases the \textit{distances} are between PDFs and then normalized following the description described in Introduction. The measurements of \textit{distance} between the two populations under investigations are observed to vary a lot - lowest being $n_0(W_1)=0.0024$ to highest being $1.0$ in multiple cases. The values of $1.0$ indicate these metrics' insensitivity between fully-disjoint and maximally-disjoint sets/functions. It was observed that Hellinger, $KS$ and $FR$ distances are most prone to this, whereas $W_1$ and $L_\infty$ distances are the least prone. But $W_1$ and $L_\infty$ distances also observed to be unstable under discretization lengths and in low statistics, as can be seen from Fig. \ref{fig:dist_vs_binsize} and Fig. \ref{fig:dist_vs_events_frac}. From Tab. \ref{tab:distance_metrics}, it also can be noted that $W_2$ and $\sqrt{JS}$ distances become saturated only under $n_3(x)$, but $W_2$ is very unstable under both discretization lengths and low statistics - which leaves $\sqrt{JS}$ to be the most reliable metric among the metrics investigated here.

\begin{figure}[!htbp]
    \centering
    \includegraphics[width=0.9\linewidth]{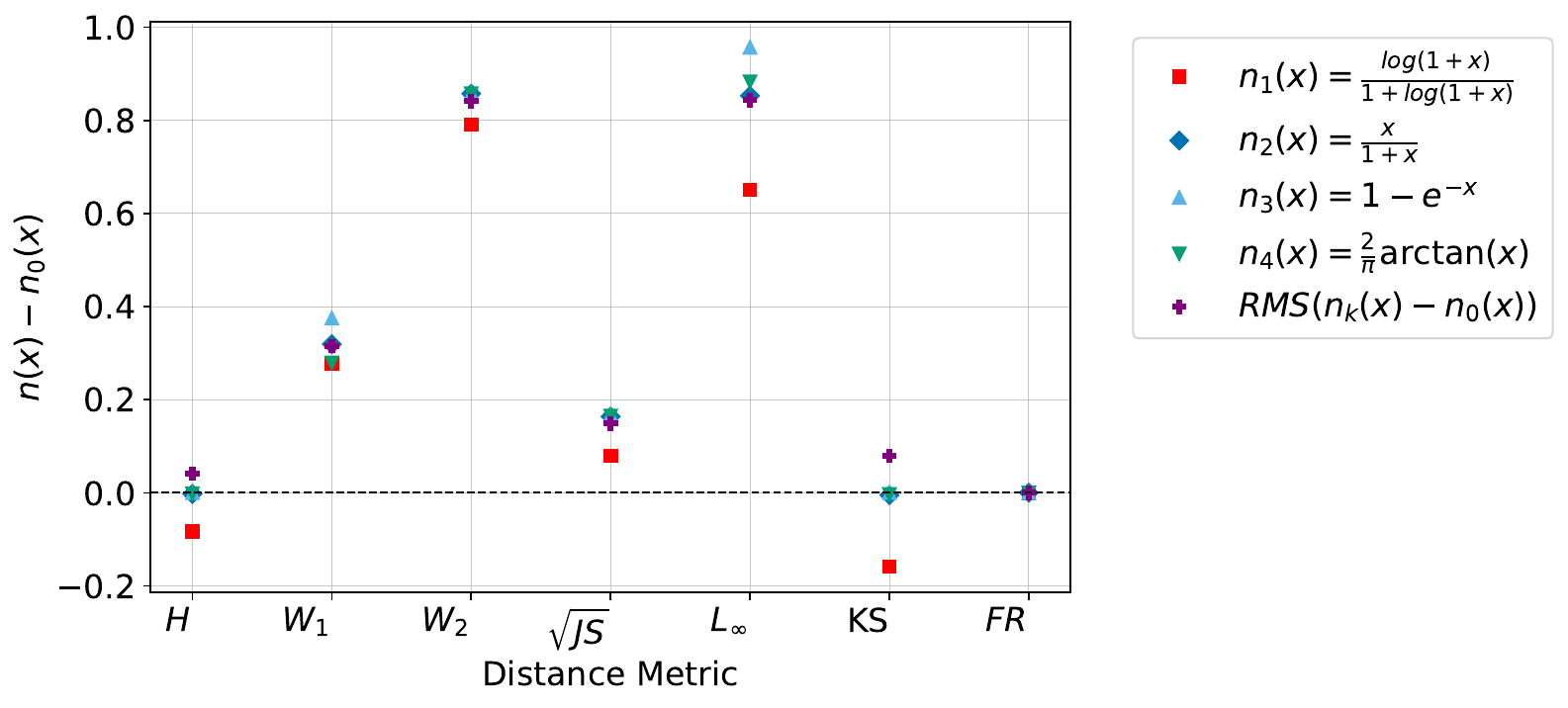}
    \caption{Effect of normalization w.r.t. no normalization, calculated by $n_k(x)-n_0(x)$. Green crosses refer to RMS values of this effect. Hellinger distance and $\sqrt{JS}$ distance are found to be the least impacted by the choice of normalizing function, whereas $L_\infty$ and $FR$ distances are found to be impacted the most}
    \label{fig:norm_comparison}
\end{figure}

\begin{figure}[!htbp]
    \centering
    \includegraphics[width=0.9\linewidth]{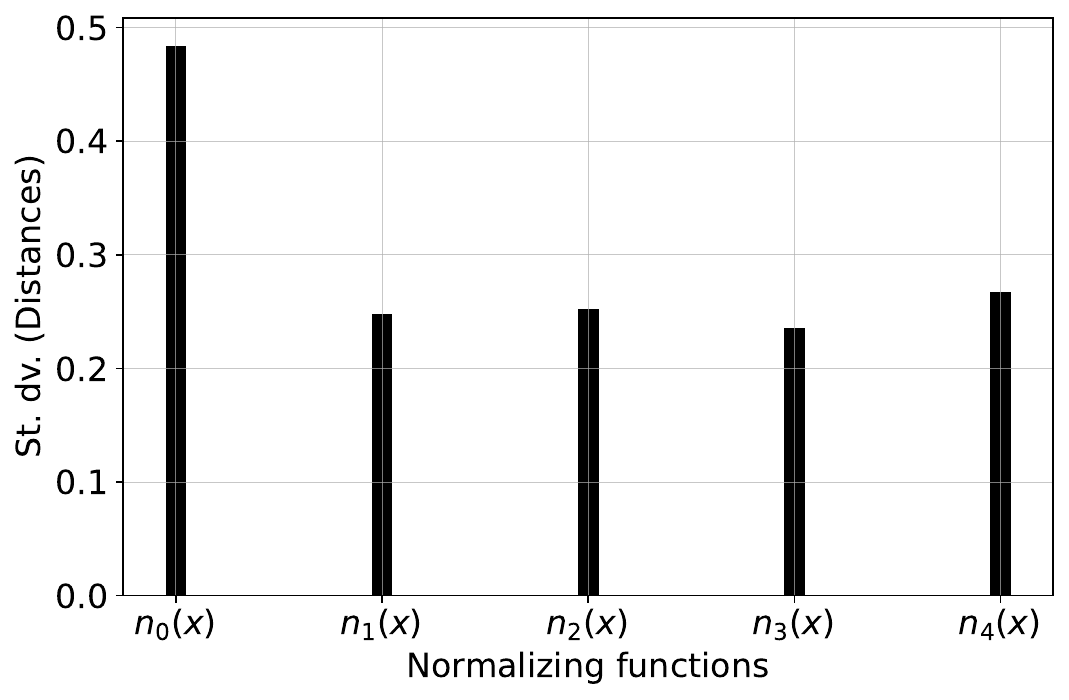}
    \caption{Standard deviations of the distances measured by all metrics under specific normalization functions. It can be observed that manually defined normalizing functions have in general lower standard deviations. No distinctions between the normalizing functions used in this analysis is evident.}
    \label{fig:stdv_comparison}
\end{figure}

It is also observed that Hellinger, $\sqrt{JS}$ and $KS$ distances are comparable when normalized and not normalized. They are same for Fisher-Rao distance and wildly different for other metrics. It is shown in Fig. \ref{fig:norm_comparison}. Fig. \ref{fig:stdv_comparison} shows the standard deviations of the measurements by all metrics under each normalization condition. It is observed that manually defined normalization functions yield lower standard deviations in general, denoting that they bring different distance metrics closer to each other compared to not normalizing.

\begin{table}[!htbp]
\centering
\begin{tabular}{|c|c|c|c|c|c|c|c|}
\hline
\textbf{} & \textbf{$H$} & \textbf{$W_1$} & \textbf{$W_2$} & \textbf{$\sqrt{}JS$} & \textbf{$L_\infty$} & \textbf{$KS$} & \textbf{$FR$} \\
\hline
$n_0(x)$ & 1.0000 & 0.0024 & 0.1419 & 0.8326 & 0.0431 & 1.0000 & 1.0000 \\
$n_1(x)$ & 0.9173 & 0.2795 & 0.9327 & 0.9119 & 0.6935 & 0.8414 & 1.0000 \\
$n_2(x)$ & 0.9975 & 0.3216 & 0.9994 & 0.9964 & 0.8959 & 0.9950 & 1.0000 \\
$n_3(x)$ & 1.0000 & 0.3775 & 1.0000 & 1.0000 & 0.9998 & 1.0000 & 1.0000 \\
$n_4(x)$ & 0.9984 & 0.2818 & 0.9996 & 0.9977 & 0.9264 & 0.9968 & 1.0000 \\
\hline
\end{tabular}
\caption{Measurements of distances between electron and photon population in the dataset. $n_0(x)$ refers to no normalization applied and $n_k(x)$ refers to the corresponding normalization functions. All of these values are dimensionless and expected to be in $[0,1)$.}
\label{tab:distance_metrics}
\end{table}

\begin{figure}[!htbp]
    \centering
    \includegraphics[width=0.9\linewidth]{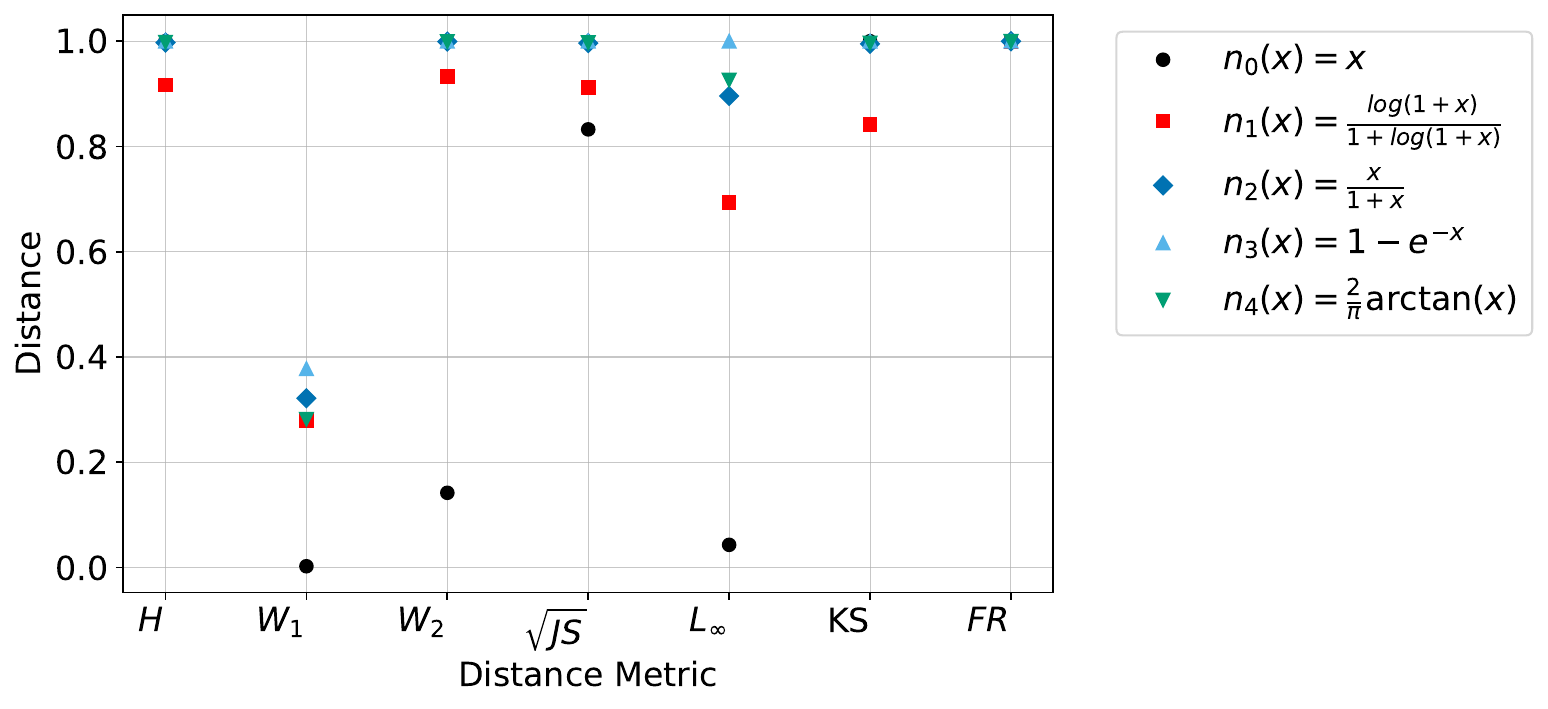}
    \caption{\textit{Distances} between electron and photon population for each metric under various normalization function. $n_0(x)$ refers to no normalization applied and $n_k(x)$ refers to the corresponding normalization function. It is a direct visualization of Tab. \ref{tab:distance_metrics}.}
    \label{fig:distance_metrics}
\end{figure}

\FloatBarrier

\section{Discussion and Conclusion}
A systematic study has been undertaken in this analysis to compare various distance metrics using electron and photon events from \iso{Kr}{83} atoms using an HPGe detector. These detectors produce distinct signatures for electrons and photons, which were parameterized in two separate ways first, then used to generate fully, but not maximally, disjoint PDF/PMFs on the said parameter space, and finally these PDF/PMFs were used to perform the analysis. Stability of these metrics were investigated under various combinations of sample sizes, discretization lengths and normalizing functions. It was found that $\sqrt{JS}$ distance is the \textit{most reliable}, defined as a combination of non-maximality preservation and stability, distance metric based on this analysis. In terms of normalizing functions, manually defined functions were observed to be slightly more robust, but no significant distinction among them were observed. It is to note that these functions can be generalized to produce different formulations, along with inclusion of more normalizing functions and distance metrics, which can be subjects to further investigations.


\begin{acknowledgments}
    We are grateful for many productive discussions and effort provided by many people at CENPA (UW) including Alejandro Garcia, Drew Byron, Brent VanDevender, Steve Elliott, David Radford, and the engineering support at CENPA including David Peterson, Tim Van Wechel, Ryan Roehnelt, Nate Miedema, Matt Kallander, and Tom Burritt. 
    We also acknowledge the efforts provided by Dr. Jason Detwiler (Professor, Department of Physics, UW) and Dr. Walter Pettus (Assistant Professor, Department of Physics, IU) through many illuminating conversations.
    This material is based upon work supported by the U.S.~Department of Energy, Office of Science, Office of Nuclear Physics under contract / award number DE-AC02-05CH11231. 
    This research used resources provided by the National Energy Research Scientific Computing Center, a U.S.~Department of Energy Office of Science User Facility.
\end{acknowledgments}
\bibliographystyle{unsrtnat}
\bibliography{references}

\end{document}